# Combined effects of anisotropy and tension-compression asymmetry on the torsional response of AZ31 Mg

Nitin Chandola[a], Ricardo A. Lebensohn[b], Benoit Revil-Baudard[a], Oana Cazacu[a*], Raja K. Mishra[c], Frédéric Barlat[d]

[a] Department of Mechanical and Aerospace Engineering, University of Florida, REEF, Shalimar, FL 32579, USA.

[b] Materials Science and Technology Division, Los Alamos National Laboratory, Los Alamos, NM 87545, USA.

[c] General Motors Research and Development Center, 30500 Mound Road, Warren, MI 40890-9055, USA.

[d] Graduate Institute of Ferrous Technology, Pohang University of Science and Technology, Republic of Korea

## Abstract

In this paper it is demonstrated that only by accounting for the combined effects of anisotropy and tension-compression asymmetry both at single crystal and polycrystal levels, it is possible to explain and accurately predict the peculiarities of the room-temperature torsional response of a strongly textured AZ31 Mg material. This is shown by using two modeling frameworks, namely: the viscoplastic self-consistent (VPSC) polycrystal model that accounts for tension-compression asymmetry of the mechanical response of a polycrystalline aggregate due to the occurrence of twinning at single crystal level, and an anisotropic plasticity model based on an orthotropic yield criterion that accounts for tension-compression asymmetry in plastic flow at macroscopic level, developed by Cazacu et al. (2006). It is shown that unlike Hill's (1948), the latter macroscopic criterion quantitatively predicts the experimental results, namely: that the sample with axial direction along the rolling direction contracts, while the sample with axial direction along the normal direction elongates. Moreover, it is demonstrated that these experimentally observed axial effects in torsion can be also quantitatively predicted with the VPSC polycrystal model only if both slip and twinning are considered active at single crystal level. On the other hand, if it is assumed that the plastic deformation is fully accommodated by crystallographic slip, the predicted axial strains are very close with that obtained with Hill (1948) criterion, which largely underestimate the measured axial strains.

Keywords: orthotropy; strength differential effects; twinning; Swift effects, Magnesium (AZ31).

[*] Corresponding author: Tel: +1 850 833 9350; fax: +1 850 833 9366.
E-mail address: cazacu@reef.ufl.edu

# 1. Introduction

The *Swift effect* refers to the observed plastic axial strains in a metallic specimen under torsion. This phenomenon was first described by Swift (1947), who reported plastic elongation of cylindrical specimens (solid rods and tubes) made of different materials with cubic crystal structure (e.g. stainless steel, aluminum, brass) and conjectured that strain hardening is the cause of these axial effects under torsional loadings. As stated by the author himself, this explanation is largely speculative and was later invalidated by Billington (1977 a-c). Hill (1950) stated that if a material is isotropic the specimen should not change its length under torsion and that Swift effect is the result of texture-induced anisotropy. This remains the prevalent view. Nowadays, it is generally accepted that the axial effects in torsion are due to texture-induced anisotropy, and therefore Swift effect is mainly modeled in the framework of crystal plasticity.

The majority of studies were devoted to materials with cubic crystal structure for which it is assumed that the plastic deformation is fully accommodated by slip. The homogenized response of the polycrystal was obtained using the upper-bound Taylor approximation (deformation gradient within each grain has a uniform value throughout the aggregate), or the intermediate viscoplastic self-consistent formulation (Lebensohn and Tome, 1993). For example, Toth et al. (1990) used a rate sensitive Taylor-type crystal plasticity model to simulate length changes and the texture developed during unconstrained shear of thin-walled tubes of pure copper. Axial effects were qualitatively predicted for very large shear strains, the simulations indicating a strong influence of the strain rate sensitivity at single crystal level. Habraken and Duchene (2004) and Duchene et al. (2007) have also confirmed the significant influence of the strain rate sensitivity on the predictions of axial effects in copper under free end torsion. However, the predicted axial strains largely underestimate the experimental ones. Some other studies have been devoted to modeling cyclic Swift effect at large strains using phenomenological models. Generally, yielding and hardening are considered anisotropic (see, for example, Kuroda (1999) who used the orthotropic Hill's (1948) yield criterion in conjunction with anisotropic hardening).

Regarding polycrystalline materials with hexagonal crystal structure (hcp), the past decade has witnessed a renewed interest in improving the fundamental understanding of their plastic behavior. In particular, a large number of experimental studies have been conducted on magnesium alloys in order to identify the crystallographic plastic deformation mechanisms at room temperature and their effects on the macroscopic response in uniaxial tension and compression, as well as in simple shear along different orientations (e.g. Barnett, 2007a-b; Jiang et al., 2008; Khan et al., 2011, etc.). In comparison, there is little information concerning the mechanical behavior of Mg and its alloys in torsion. Free-end torsion tests on solid bars of pure Mg and Mg alloy AZ71 with the axis along the rolling direction at room temperature, $150^0$ C, and $250^0$ C were reported by Beausir et al. (2009) and Biswas et al. (2013). While for both materials the shear deformation was very limited at room temperature, at 250°C the maximum plastic shear was much larger for the pure Mg ($\gamma =1.7$ as compared to ~0.2 at room temperature). For all temperatures, shortening of the specimens was observed. In the simulations, deformation twinning was not considered for both materials because only the high-temperature behavior (250 C) was modeled. Very recently, Guo et al. (2013) reported room-temperature free-end torsion tests on AZ31 Mg alloy cylindrical solid bars machined from a rolled plate with a very strong initial basal texture. Specimens with longitudinal axis oriented along the rolling direction and the through-thickness (normal direction) of the plate were tested. It was found that while the sample with axial direction along the rolling direction contracts, the sample with axial direction along the normal direction elongates. These axial effects were attributed to tensile twinning, which, in turn, induces texture evolution in the material. For α-titanium tubes with basal texture deforming by slip and twinning, systematic elongation of the specimens is observed irrespective of the orientation of the long axis of the specimen as evidenced by the analysis of tubes obtained by flow-forming (see Scutti, 2000).

In summary, initial anisotropy and texture-induced anisotropy associated to either slip or deformation twinning has been considered the only causes of Swift effect in polycrystalline metals. However, Billington (1977) reported that there are metals with cubic crystal structure that display significant Swift effect but remain isotropic over the entire range of plastic deformation. Very recently, a new interpretation of this phenomenon in isotropic materials was provided by Cazacu et al. (2013) for monotonic loadings and by Cazacu et al. (2014) for cyclic loadings using a macroscopic modeling framework. Analytical calculations using Cazacu et al. (2006) criterion showed that the occurrence of axial effects is related to a slight difference between the uniaxial yield in tension and compression of the given isotropic material. More specifically, if the uniaxial yield stress in tension is larger than that in compression, the specimen will contract when twisted, while, if the yield stress in uniaxial compression is larger than that in uniaxial tension, it will elongate.

In this paper it is demonstrated that only by accounting for tension-compression asymmetry in plastic flow it is possible to explain and predict with accuracy the peculiar features of the room-temperature torsional response of AZ31 Mg. This is established using two approaches: (i) the polycrystalline VPSC homogenization approach (Lebensohn and Tomé, 1993), which accounts for the asymmetry in plastic flow at polycrystal level induced by that at single-crystal level associated with deformation twinning, and (ii) an elastic/plastic approach based on an orthotropic yield criterion that accounts for tension-compression asymmetry in plastic flow at macroscopic level, developed by Cazacu et al. (2006). These models are presented in Section 2 and further applied to the description of the mechanical response of AZ31 Mg. The VPSC model is calibrated based on uniaxial tension and compression data, and further used to simulate simple shear and free-end torsion (Section 3). For the first time, it is shown that although the stress vs. shear strain response can be described considering only prismatic $\{1\bar{1}00\}<11\bar{2}0>$, basal $\{0001\}<1\bar{2}10>$ and pyramidal <c+a> $\{10\bar{1}1\}<\bar{1}\bar{1}23>$ slip modes to be active, the axial strains and texture evolution can be predicted with accuracy only if both pyramidal <a> $\{10\bar{1}1\}<\bar{1}2\bar{1}0>$ slip and tensile twinning $\{10\bar{1}2\}<\bar{1}101>$ are also considered operational. Furthermore, the agreement with the experimental results and the level of accuracy of the predictions of shear and axial strains in free-end torsion (i.e. Swift effect) is similar to that of Cazacu et al. (2006) criterion. For completeness, the torsional response of AZ31 Mg is also simulated using Hill's (1948) orthotropic yield criterion that neglects tension-compression asymmetry in the plastic flow. It is shown that Hill's criterion drastically overestimates the axial strains that occur in torsion in the rolling direction and cannot capture at all axial effects that develop in the normal direction torsion. Interestingly, it is shown that if it is considered that plastic deformation is accommodated only by crystallographic slip, the VPSC predictions are very close with that according to Hill (1948), the discrepancy between VPSC and experimental data being of the same order.

Regarding notations, vectors and tensors are denoted by boldface characters. If **A** and **B** are second-order tensors, the contracted tensor product between such tensors is defined as: $\mathbf{A}:\mathbf{B} = A_{ij}B_{ij}$, i, j = 1…3; if **u** and **v** are two vectors, their dyadic product is the second rank tensor $(\mathbf{u} \otimes \mathbf{v})_{ij} = u_i v_j$.

## 2. Constitutive models

### 2.1. Polycrystal model

The VPSC model will be used to gain insights into the role of specific single-crystal plastic deformation mechanisms on the macroscopic response of AZ31 Mg. This model is only briefly presented in what follows (a detailed description can be found in the review article by Tomé and Lebensohn (2004)). The polycrystal is represented by a finite set of orientations, each one representing a given volume fraction chosen to reproduce the initial texture. The total deformation of the polycrystal is obtained by imposing successive strain increments and calculating the resulting strains in the grains. The grain reorientations associated with these plastic strains lead to texture evolution. A self-consistent approach is used to model the interaction of a grain with its surroundings. Each grain is treated as an anisotropic, viscoplastic, ellipsoidal inclusion embedded in a uniform matrix having the unknown properties (to be determined) of the polycrystal. Elastic deformations are neglected. Each deformation system (s) is characterized by a vector $\mathbf{n}^s$ (normal to the slip or twinning plane) and a vector $\mathbf{b}^s$ (Burgers vector or twinning shear direction). The local constitutive behavior (at the grain level) is described by:

$$\mathbf{d}^g = \sum_s \mathbf{m}^s \dot{\gamma}^s = \dot{\gamma}_0 \sum_s \mathbf{m}^s \left( \frac{\left| \mathbf{m}^s : \boldsymbol{\sigma}^g \right|}{\tau_c^s} \right)^n \times \mathrm{sgn}\left( \mathbf{m}^s : \boldsymbol{\sigma}^g \right), \qquad (1)$$

where $\mathbf{m}^s = \frac{1}{2}(\mathbf{b}^s \otimes \mathbf{n}^s + \mathbf{n}^s \otimes \mathbf{b}^s)$ and $\dot{\gamma}^s$ are the Schmid tensor and the shear rate of system (s), $\mathbf{d}^g$ and $\mathbf{\sigma}^g$ are the local averages of the strain rate and stress fields in grain (g), $\dot{\gamma}_0$ is a reference shear rate and *n* is a rate sensitivity parameter. Equation (1) expresses that the deformation rate is given by the sum over all the shear rates contributed by all systems. For both slip and twinning, the activation criterion is given by the expression in parenthesis, which expresses that the activity on each deformation system (s) increases when the resolved shear on that system (given by $\mathbf{m}^s : \mathbf{\sigma}^g$) approaches a threshold value $\tau_c^s$. Strain-hardening is incorporated by allowing the threshold stress $\tau_c^s$ to evolve according to:

$$\tau_c^s = \tau_0^s + (\tau_1^s + \theta_1^s \Gamma)\left(1 - \exp\left(-\frac{\theta_0 \Gamma}{\tau_1^s}\right)\right), \qquad (2)$$

where $\tau_0^s$, $\tau_1^s$, $\theta_0^s$ and $\theta_1^s$ are constants, and $\Gamma = \sum_s \dot{\gamma}^s \Delta t$ is the accumulated shear in all active deformation systems. In addition, it is possible to incorporate self and latent hardening. More specifically, the increase in the threshold stress is calculated as:

$$\Delta \tau_o^{s(r)} = \frac{d\tau^{*s(r)}}{d\Gamma^{(r)}} \sum_{k'} h^{ss'} \dot{\gamma}^{s'(r)} \Delta t, \qquad (3)$$

where the coefficients $h^{ss'}$ empirically account for the obstacles that new dislocations (or twins) associated with system *s'* create for the propagation of dislocations (or twins) on system *s*. Following the approach proposed by Van Houtte (1978), twinning is treated as a pseudo-slip mechanism. Specifically, it differs from slip in its directionality, which means that activation of twinning is allowed only if the resolved shear stress is positive. The twinning contribution to texture development is accounted for by means of the so-called Predominant Twin Reorientation (PTR) scheme (Tomé et al, 1991). In this scheme, the grains where twinning is most active are determined. The fixed number of orientations that represent the polycrystal is maintained throughout the deformation history. The volumetric effect of twinning reorientation on texture development is modeled by reorienting some of these grains completely into the orientation of their most active twinning systems.

**2.2. Macroscopic elastic/plastic model**

To explain and model Swift effect in AZ31 Mg, an elastic-plastic modeling approach in the framework of the mathematical theory of plasticity is also considered. The yield condition is based either on the orthotropic yield function of Hill (1948) or on that of Cazacu et al. (2006), and for both cases hardening is considered isotropic. The general form of the governing equations are first presented and followed by the specific expressions of these yield criteria.

Tensile strains and stresses are considered positive. The total rate of deformation **D** is the sum of the elastic part (**D**-**D**$^P$), and the plastic part, **D**$^P$. The elastic stress-strain relationship is given by

$$\dot{\boldsymbol{\sigma}} = \mathbf{C}^e : (\mathbf{D} - \mathbf{D}^p), \tag{4}$$

where $\dot{\boldsymbol{\sigma}}$ is the Green-Naghdi rate, which is an objective rate of the Cauchy stress tensor $\boldsymbol{\sigma}$ (see Green and Naghdi, 1965; ABAQUS, 2009). In this work we assume linear-elastic isotropy, so with respect to any Cartesian coordinate system, $\mathbf{C}^e$ is expressed as

$$C^e_{ijkl} = 2G\,\delta_{ik}\delta_{jl} - \left(K - \frac{2}{3}G\right)\delta_{ij}\delta_{kl}, \tag{5}$$

where i, j, k, l = 1...3, $\delta_{ij}$ is the Kronecker delta tensor, and G and K are the shear and bulk moduli, respectively. The rate of deformation tensor is given by the associated flow rule:

$$\mathbf{D}^p = \dot{\lambda}\frac{\partial F}{\partial \boldsymbol{\sigma}}, \tag{6}$$

where F is the yield function and $\dot{\lambda} > 0$ is the plastic multiplier. Strain hardening is considered isotropic and governed by the accumulated plastic strain. Thus, the plastic potential F in Eq. (6) is of the general form:

$$F(\boldsymbol{\sigma}, \bar{\varepsilon}^p) = \bar{\sigma}(\boldsymbol{\sigma}, \bar{\varepsilon}^p) - Y(\bar{\varepsilon}^p), \tag{7}$$

where $\bar{\sigma}$ is the effective stress associated with the given yield criterion, $\bar{\varepsilon}^p$ is the equivalent plastic strain which is calculated based on the plastic work equivalence principle (i.e. it is the work-conjugate of $\bar{\sigma}$), while $Y = Y(\bar{\varepsilon}_p)$ is the hardening law. Specifically, a Voce-type hardening law is considered:

$$Y(\bar{\varepsilon}^p) = A_0 - A_1 \exp(-A_2 \bar{\varepsilon}^p), \tag{8}$$

where $A_0$, $A_1$, $A_2$ are constants.

### 2.2.1 Cazacu et al. (2006) yield criterion

Cazacu et al. (2006) proposed an anisotropic yield criterion that accounts for both plastic anisotropy and tension-compression asymmetry, which is particularly suitable for modeling hcp materials. The effective stress $\bar{\sigma}$ associated with this criterion (denoted CPB06) is:

$$\bar{\sigma} = B\left[\left(|\Sigma_1| - k\Sigma_1\right)^a + \left(|\Sigma_2| - k\Sigma_2\right)^a + \left(|\Sigma_3| - k\Sigma_3\right)^a\right]^{1/a}, \tag{9}$$

where $k$ is a material parameter accounting for the tension-compression asymmetry in yielding, $a$ is a homogeneity constant (usually, a = 2). In Eq. (9) $\Sigma_1, \Sigma_2, \Sigma_3$ are the principal values of the transformed stress $\boldsymbol{\Sigma}$ defined as:

$$\boldsymbol{\Sigma} = \mathbf{C} : \boldsymbol{\sigma}' \tag{10}$$

where $\boldsymbol{\sigma}'$ is the deviator of the Cauchy stress tensor $\boldsymbol{\sigma}$, and $\mathbf{L}$ is a fourth-order symmetric tensor characterizing the plastic anisotropy of the material. For example, for an orthotropic material, in the coordinate system associated with the material symmetry axes (*x*, *y*, *z*) (for a plate material these axes of symmetry are the rolling (RD), transverse (TD), and normal direction (ND) or through-thickness plate direction, respectively) the transformed stress tensor $\boldsymbol{\Sigma}$ is given by:

$$\begin{bmatrix} \Sigma_{xx} \\ \Sigma_{yy} \\ \Sigma_{zz} \\ \Sigma_{xy} \\ \Sigma_{yz} \\ \Sigma_{xz} \end{bmatrix} = \begin{bmatrix} C_{11} & C_{12} & C_{13} & 0 & 0 & 0 \\ C_{12} & C_{22} & C_{23} & 0 & 0 & 0 \\ C_{13} & C_{23} & C_{33} & 0 & 0 & 0 \\ 0 & 0 & 0 & C_{44} & 0 & 0 \\ 0 & 0 & 0 & 0 & C_{55} & 0 \\ 0 & 0 & 0 & 0 & 0 & C_{66} \end{bmatrix} \begin{bmatrix} \sigma'_{xx} \\ \sigma'_{yy} \\ \sigma'_{zz} \\ \sigma'_{xy} \\ \sigma'_{yz} \\ \sigma'_{xz} \end{bmatrix} \qquad (11)$$

In Eq. (9) B is a constant defined such that $\bar{\sigma}$ reduces to the tensile yield stress in the RD direction or **x**, i.e.

$$B = 1 / \left[ \left( |\Phi_1| - k\Phi_1 \right)^a + \left( |\Phi_2| - k\Phi_2 \right)^a + \left( |\Phi_3| - k\Phi_3 \right)^a \right]^{1/a} \qquad (12)$$

where $\Phi_1 = (2C_{11} - C_{12} - C_{13})/3$, $\Phi_2 = (2C_{12} - C_{22} - C_{23})/3$; $\Phi_3 = (2C_{13} - C_{23} - C_{33})/3$.

**2.2.2 Hill (1948) orthotropic yield criterion**

For comparison purposes, Hill's (1948) orthotropic yield criterion is also used to describe yielding of AZ31 Mg. With respect to the axes of orthotropy (*x*, *y*, *z*), the equivalent stress associated to this criterion is expressed as:

$$\bar{\sigma}_{Hill} = \sqrt{F(\sigma_{yy} - \sigma_{zz})^2 + G(\sigma_{zz} - \sigma_{xx})^2 + H(\sigma_{xx} - \sigma_{yy})^2 + 2L(\sigma_{yz}^2) + 2M(\sigma_{zx}^2) + 2N(\sigma_{xy}^2)} \qquad (13)$$

where F, G, H, L, M and N are material parameters (anisotropy coefficients).

## 3. Prediction of the mechanical response of AZ31 Mg in free-end torsion using the polycrystal model

### 3.1 Identification of VPSC material parameters

The polycrystal model VPSC has been previously used to simulate the plastic deformation of AZ31 Mg at room temperature. The focus of these studies was on modeling the deformation behavior under uniaxial tension or compression (e.g. Proust et al., 2009; Wang et al., 2010, etc.). Concerning simulation of the shear response, most of the efforts have been focused on understanding and modeling the high-temperature behavior of Mg and its alloys. It was established that at high temperature twinning activity is negligible, the plastic deformation being accommodated by pyramidal <a> and pyramidal <c+a> slip (e.g. for simulation of pure Mg, see Agnew et al. 2005; for pure Mg and AZ71 Mg, see Beausir et al. 2009).

To the best of our knowledge, modeling of the microstructure evolution and its effects on the macroscopic stress-strain response in shear of AZ31 Mg at room temperature were not reported. One of the objectives of this paper is to understand which are the single crystal plastic deformation mechanisms responsible for the experimentally observed texture evolution in shear and ultimately contribute to an improved understanding of room temperature Swift effect in this material.

Identification of the material parameters of VPSC is based on the room-temperature quasi-static data (strain rate of 1/s) reported by Khan et al. (2011) on AZ31 (3wt%Al, 1wt%Zn, Mg bal.). The uniaxial tensile and compression tests on specimens cut along the RD, TD, and 45° to the rolling direction of the sheet have revealed that the plastic response of this material is highly directional. The strong difference between the yield in tension and compression at the onset of plastic deformation (0.002 strain) and the unusual hardening behavior in uniaxial compression in RD (S-shape sigmoidal appearance of the stress-strain curve, see also Fig. 2) was attributed to the occurrence of tensile twinning (similar data and interpretation of the results have been also reported by Lou et al., 2007).

In simple shear along RD direction, Khan et al. (2011) reported that the volume fraction of tensile twins at 20% equivalent strain is of 0.506. The shear specimen has a higher volume fraction of twins as compared to the tension specimens, but much lower than in compression (80% at the same level of equivalent plastic strain).

Since in a typical polycrystal plasticity simulation, the polycrystal is represented with a discrete set of grains, in order to correctly capture the initial texture of the material it is necessary to consider a sufficiently large number of grains. Thus, we used the entire EBSD scan (784x808 μm area) to generate 2762 weighted orientations. The corresponding initial texture used in the simulations (Fig. 1b) is similar to that reported by Khan et al. (Fig. 1a). Note that the material displays a strong basal texture with an almost equal fraction of grains having their c-axes slightly tilted away (around ±30$^0$) from the sheet normal towards +RD and –RD, respectively.

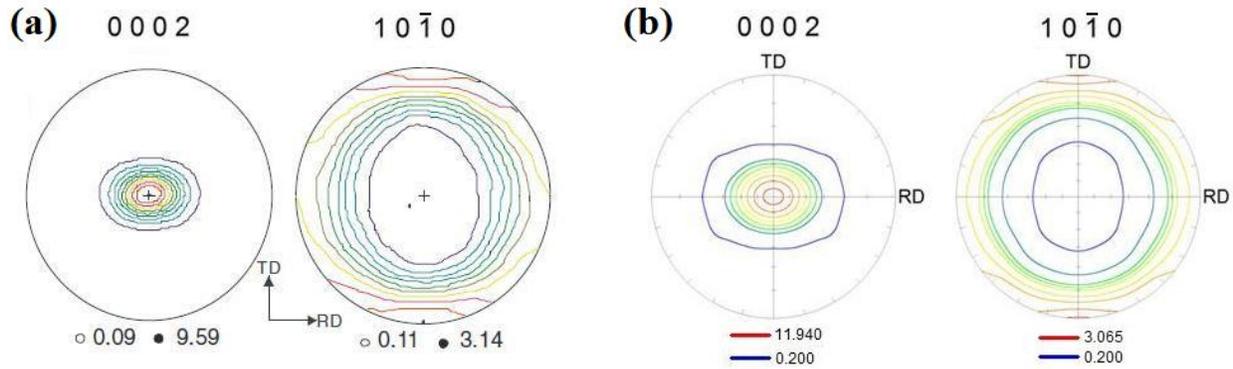

**Fig. 1**. Pole figures showing the initial texture of the AZ31 Mg sheet: (a) reported in Khan et al. (2011); (b) Measured from the entire EBSD scan and used as input for the VPSC polycrystalline model.

Based on the experimental evidence provided by Khan et al. (2011), it can be assumed that basal, prismatic, and pyramidal <c+a> slip systems as well as tensile twinning are potentially active. For these deformation systems, the set of material parameters that characterize the plastic deformation at grain level was obtained using a step-by-step procedure. Specifically, the parameters associated with basal and prismatic slip systems were identified based on the stress-strain mechanical response in uniaxial tension along RD. Next, using the uniaxial compression stress-strain data in the ND direction, the parameters associated with the pyramidal <c+a> slip system were calibrated. Finally, from the uniaxial compression test along RD, the parameters associated to the tensile twinning system were determined. In addition to these four modes, the pyramidal <a> slip mode was also considered to be active in simple shear. The hardening parameters associated with this system were calibrated based on the reported shear stress vs. shear strain curve in the RD direction. The values of the parameters associated to all five deformation modes are given in Table 1. The reference shear rate, $\dot{\gamma}_0$, and the rate-sensitivity parameter $n$ (see Eq. (1)) were prescribed to be the same for all deformation systems: $\dot{\gamma}_0 = 0.001 s^{-1}$ and $n=20$. The latent hardening coefficients $h^{ss'}$ (see Eq. 3) associated with the interaction between basal slip and tensile twinning, and respectively non-basal <a> slip and tensile twinning were set to a value of 2. On the other hand, the hardening coefficient $h^{ss'}$ associated with interaction between pyramidal <c+a> slip and tensile twinning was set to unity in order to capture the very little twinning activity observed in ND compression.

| Mode | $\tau_{0}$ [MPa] | $\tau_{1}$ [MPa] | $\theta_0$ | $\theta_1$ |
|---|---|---|---|---|
| Basal $\{0001\}<1\bar{2}10>$ | 17.5 | 5 | 3000 | 35 |
| Prismatic $\{1\bar{1}00\}<11\bar{2}0>$ | 85 | 33 | 550 | 70 |
| Pyramidal <a> $\{10\bar{1}1\}<\bar{1}2\bar{1}0>$ | 100 | 30 | 30 | 10 |
| Pyramidal <c+a> $\{10\bar{1}1\}<\bar{1}\bar{1}23>$ | 148 | 50 | 8500 | 0 |
| Tensile Twinning $\{10\bar{1}2\}<\bar{1}101>$ | 52 | 0 | 0 | 0 |

**Table 1:** Material parameters involved in the hardening laws of various deformation modes for AZ31 Mg.

Comparison between the polycrystalline model predictions of the stress-strain response in uniaxial tension and compression along the axes of orthotropy of the material (solid line) and all available data (symbols) are shown in Fig. 2. It can be concluded that the model describes very well both the strong anisotropy and the tension-compression asymmetry of the material for uniaxial loading in each orientation.

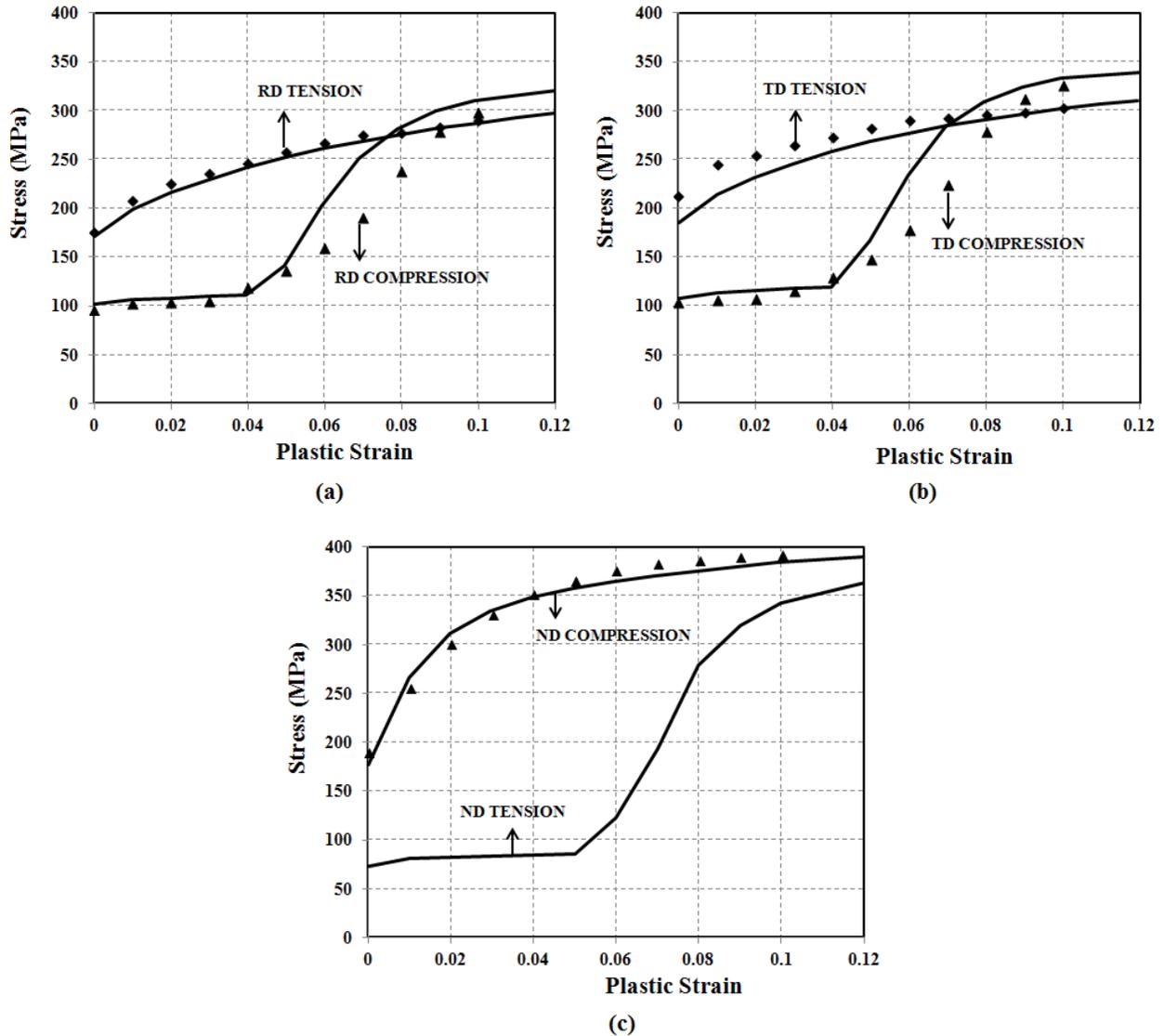

**Fig. 2**. Anisotropy of the stress-strain response in uniaxial tension and compression predicted with the VPSC model (lines) in comparison with data (symbols): along (a) Rolling Direction (RD), (b) Transverse Direction (TD), and (c) Normal Direction (ND). Data after Khan et al. (2011).

Next, we have used the calibrated VPSC model to predict the material's mechanical response for strain paths that were not considered in the model calibration. In particular, we assess the model capabilities to predict Swift effect in free-end torsion. As it is generally considered (e.g. Beausir et al., 2009) this strain path can be approximated with simple shear. We begin by simulating simple shear of a tube with the long-axis along the RD (**x**-direction). Thus, in the local coordinate Cartesian reference system, the imposed velocity gradient is:

$$\mathbf{L} = \begin{pmatrix} ? & 0 & ? \\ \dot{\gamma}_{21} & ? & ? \\ 0 & 0 & ? \end{pmatrix}_{(x,y,z)} \quad (14)$$

The calculations were terminated when the shear strain reached the value of $\gamma = 0.2$. All five deformation modes were considered to be active. Figure 3a shows a comparison between the predicted shear stress vs. shear strain response according to the polycrystal model (solid line), and mechanical data reported in Khan et al. (2011), along with the predicted textures corresponding to different levels of the shear strain, and the predicted evolution of the twin volume fraction (inset). Note that the agreement between the model and data is very good, both in terms of the macroscopic shear stress vs. shear strain response and final value of the twin volume fraction, reported in Khan et al. (2011). The predicted relative slip/twinning activities are shown in Fig. 3b. The activity plot shows that tensile twinning activity is highest at approximately 0.035 von Mises equivalent strain ($\gamma/\sqrt{3}$). At this strain level, there is a slight kink in the macroscopic stress-strain response (Fig. 3a), indicating a change in the strain-hardening rate. It is important to note that the kink in the stress-strain curve was also observed experimentally at approximately 0.025 equivalent strain (see also Lou et al. (2007)), which was attributed to tensile twinning. Comparison between the only available measured texture and that predicted by the model is shown in Fig. 4. The final texture observed in mechanical tests in RD shear is very close to that predicted by VPSC.

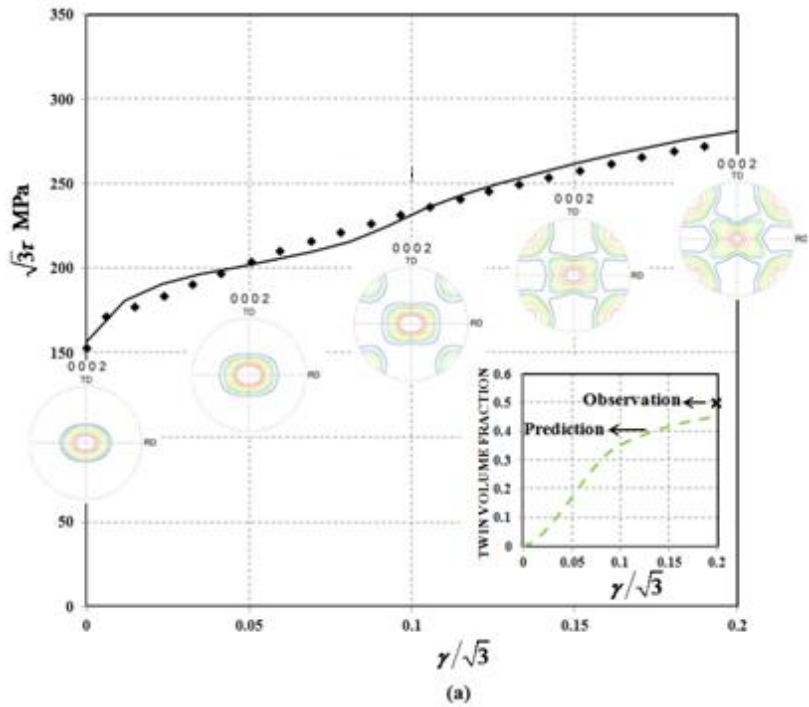

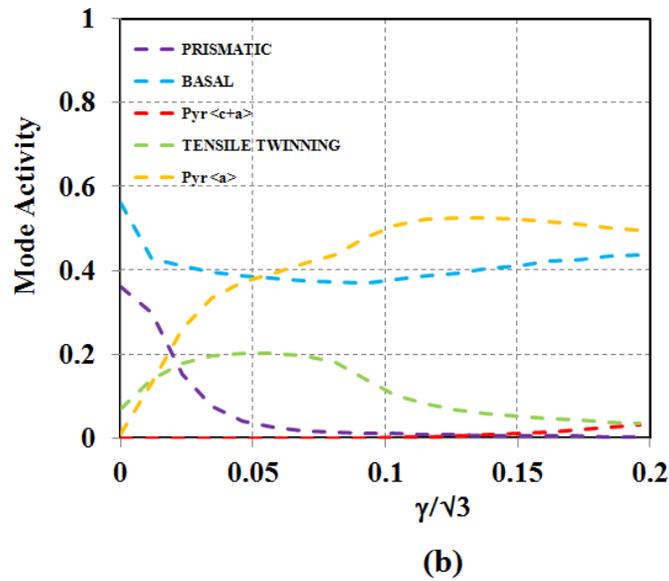

**Fig. 3**. (a) Predicted stress-strain, texture using the polycrystalline VPSC model for RD shear. Inset shows the predicted twin volume fraction evolution in RD shear; and the only available experimental measurement (symbol (x) reported by Khan et al. (2011)). (b) Predicted relative activities of each deformation mode contributing to plastic deformation.

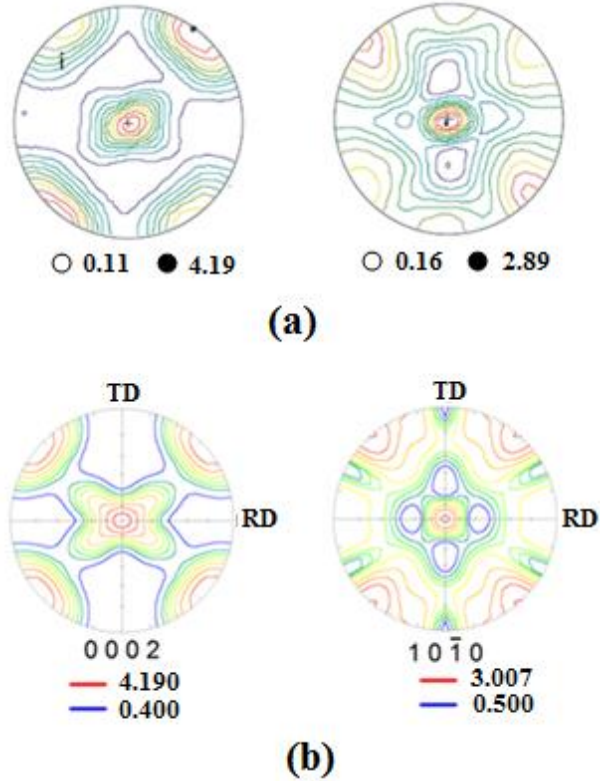

**Fig. 4:** Pole figures of AZ31 Mg sheet in RD shear test (at $\gamma/\sqrt{3} = 20\%$ strain) (a) Measured by Khan et al. 2011 (top) (b) Predicted by the VPSC model (bottom).

To gain understanding of the role played by individual plastic deformation mechanisms, two additional simulations were conducted. More specifically, simulation of the shear response was also done considering that plastic deformation is accommodated only by the four slip modes, i.e. neglecting tensile twinning activity. The predicted macroscopic response and evolving microstructure is given in Fig. 5b. It is worth noting that up to now in all the studies devoted to modeling the deformation response of AZ31 Mg using the VPSC model (Jain and Agnew, 2007; Wang et al., 2010; Guo et al.. 2013) only basal, prismatic, and pyramidal <c+a> slip modes we considered active. However, pyramidal <a> slip mode appears to be necessary to explain microstructure evolution in simple shear. In order to understand the role played by pyramidal <a> slip, simulation of the shear response was conducted considering that plastic deformation is accommodated only by basal, prismatic, and pyramidal <c+a> slip (i.e. both pyramidal <a> slip and tensile twinning was neglected, see Fig. 5c). If twining activity is neglected, the strains are slightly underestimated for $\gamma > 0.173$ (see Fig. 5b); if both tensile twining and pyramidal <a>

slip are neglected, the strains are slightly overestimated for the entire strain path history (see Fig. 5c). While this indicates that the overall agreement between simulated and experimental shear stress vs. shear strain response is good irrespective of which plastic deformation mechanisms are considered to be operational, it turns out that when twinning and pyramidal <a> slip are neglected, texture evolution is not correctly captured.

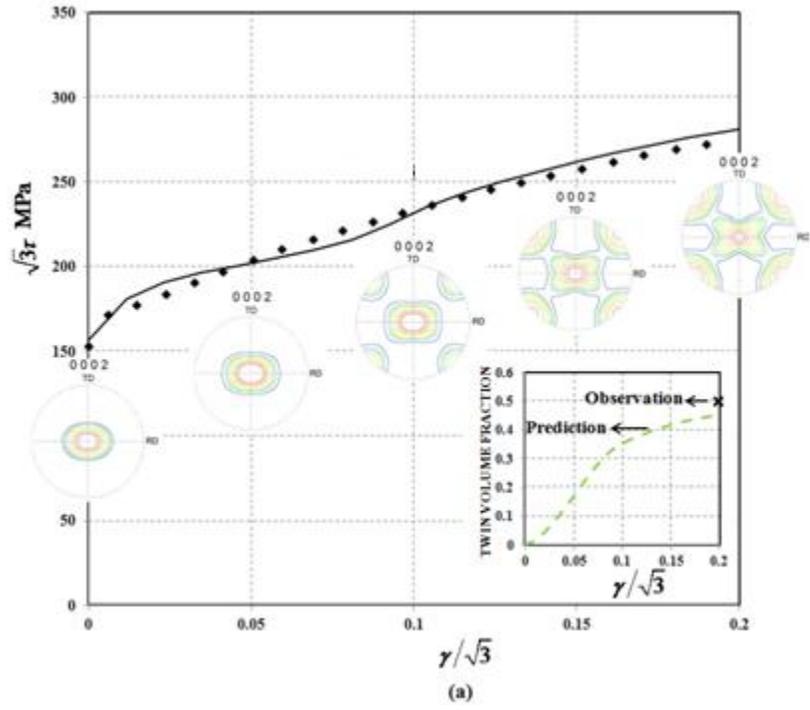

(a)

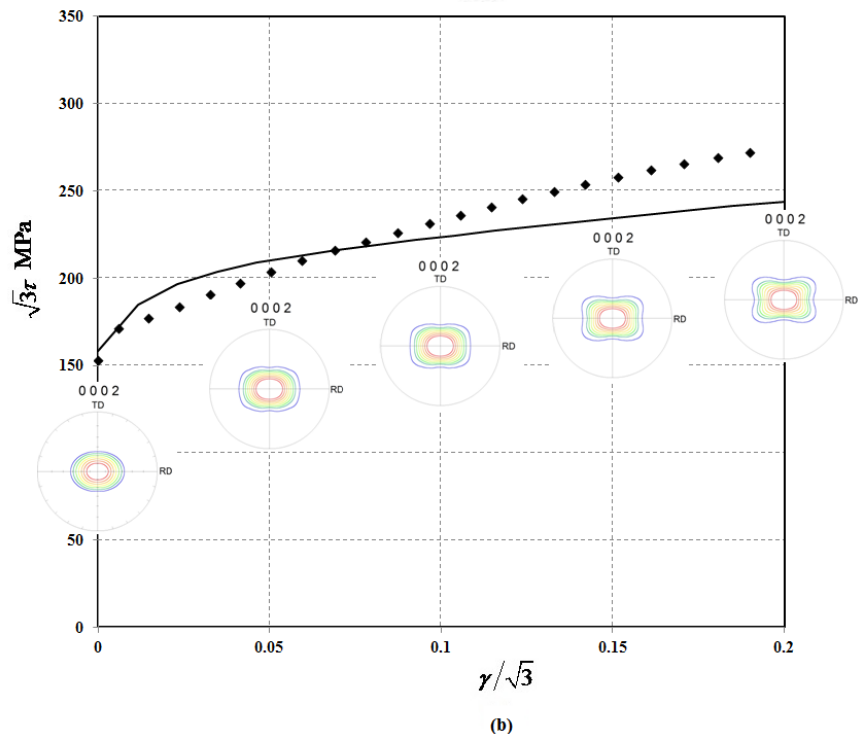

(b)

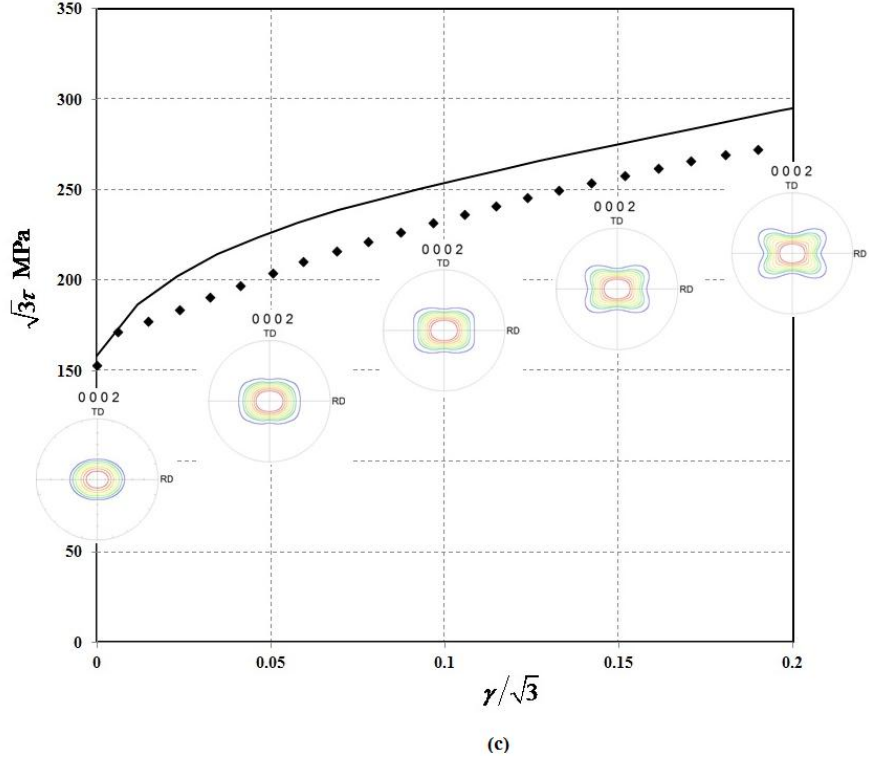

**Fig.5:** Predicted effective stress vs. effective strain response and evolution of the microstructure in RD shear using the VPSC model (solid line): (a) tensile twinning and all four slip modes considered initially active; (b) all slip modes considered active and no twinning;(c) three slip modes considered active while pyramidal <a> and tensile twinning neglected.

To demonstrate this point, the experimental and predicted final textures in each case are compared in Fig. 6. Indeed, when tensile twinning and all the four slip modes (basal, prismatic, pyramidal <a> and pyramidal <c+a> slip) are considered active, the final observed and predicted textures match extremely well. However, if twinning is not operating, it is predicted that only a rotation of the initial texture occurs when the material is subjected to simple shear. More specifically, the polycrystal model predicts that the basal pole intensity is elongated 45° away from the shear direction (which is the RD direction), and the experimentally observed rotation of the <c>-axes is not captured at all. Furthermore, comparison between Fig. 6c and Fig 6d, shows the specific role played by pyramidal <a> slip, i.e. inhibiting the rotation of the basal planes along the two mutually perpendicular directions at 45° to RD.

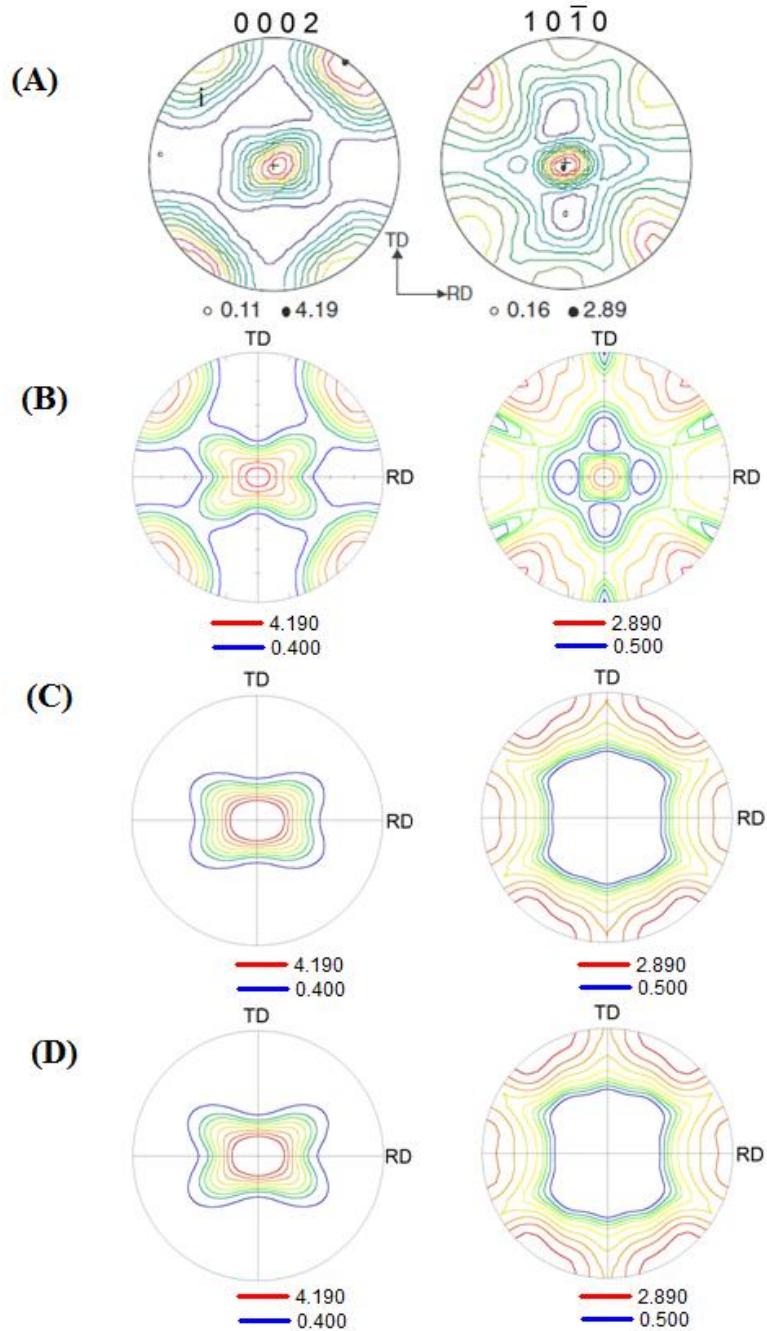

**Fig. 6:** Final texture in simple shear (at 20% strain) along RD direction obtained from: (a) Experiments; (b) VPSC model predictions assuming that all four slip modes and tensile twinning are active; (c) VPSC model predictions assuming that all slip modes are active and no tensile twinning activity; (d) VPSC predictions assuming that only three slip modes active, i.e. pyramidal <a> and tensile twinning activities neglected.

Moreover, Fig. 7 shows a comparison between the axial strain vs. shear strain predicted by the VPSC model and data on free-end torsion of an AZ31 Mg tube reported in Guo et al. (2013). As already mentioned, the applied boundary conditions in the VPSC simulations (see Eq. 14) are representative of these in free-end torsion of a tube. It is important to note that irrespective of the deformation modes that are considered to be active, shortening (i.e. axial strains negative) of the specimen along the direction of twist (RD) is qualitatively predicted. However, only in the case when tensile twinning is considered active, there is quantitative agreement with the data, in particular the slope of the experimental curve is correctly predicted. In conclusion, it has been demonstrated that only by consideration of all slip modes and tensile twinning it is possible to accurately predict both the shear stress vs. shear strain response, the axial strain vs. shear strain, and the microstructure evolution (twin volume fraction; final texture) in AZ31 Mg.

In the ND direction, only results of torsion tests on AZ31 Mg tubes reported in Guo et al. (2013) were available. These test results indicate that elongation of the specimen occurs in the ND direction. However, in the absence of relevant metallographic information (such as information about the variation in texture along the thickness of the plate, twin volume fraction measurements) it is not possible to establish, on the basis of crystal plasticity simulations alone, the role of individual plastic deformation mechanisms in simple shear along the ND direction. Therefore, VPSC simulations were conducted assuming that tensile twinning and all four slip modes are active, with constitutive parameters given in Table 1. Note that the polycrystalline model predicts correctly the nature of axial strains that develop, i.e. that in ND torsion the axial strains are positive (elongation of the sample, see Fig. 8) while the RD specimen shortens (axial strains negative, see Fig. 8).

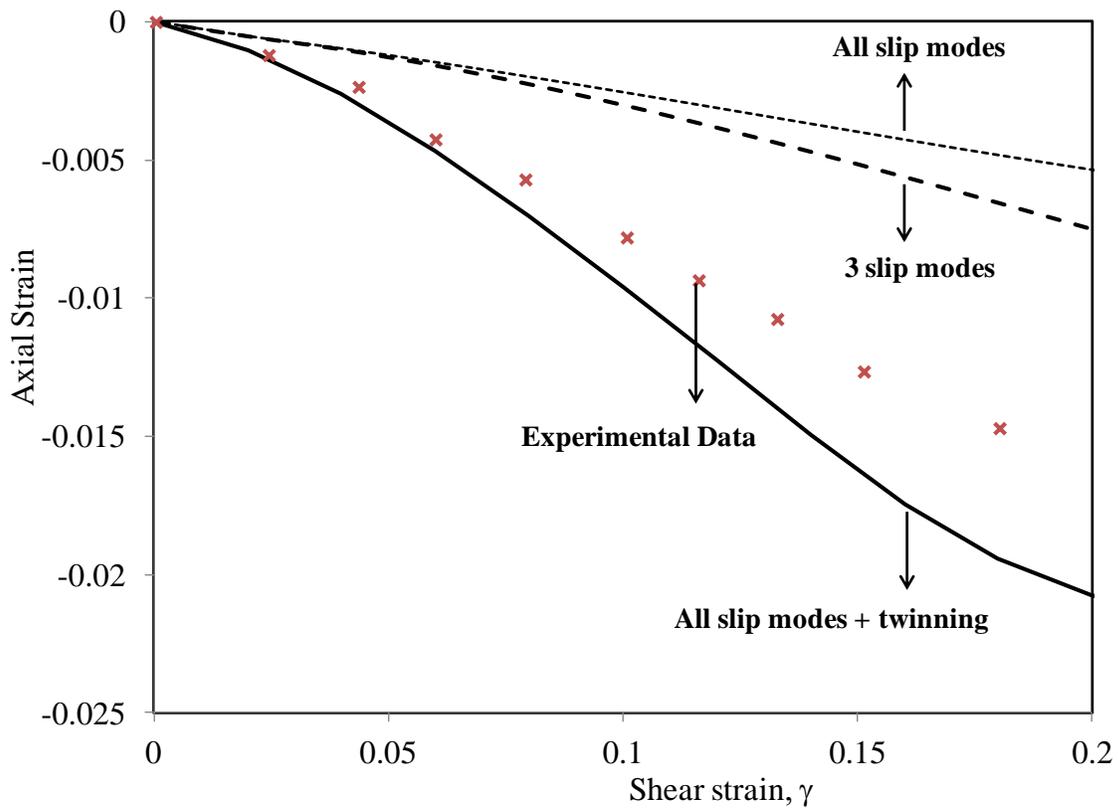

**Fig. 7:** Comparison of the axial strain vs. shear strain observed experimentally in shear along RD (symbols) and according to the VPSC model (lines). Note that only when all slip modes and tensile twinning are active, the Swift effect (axial contraction) are well described. Data from Guo et al (2013).

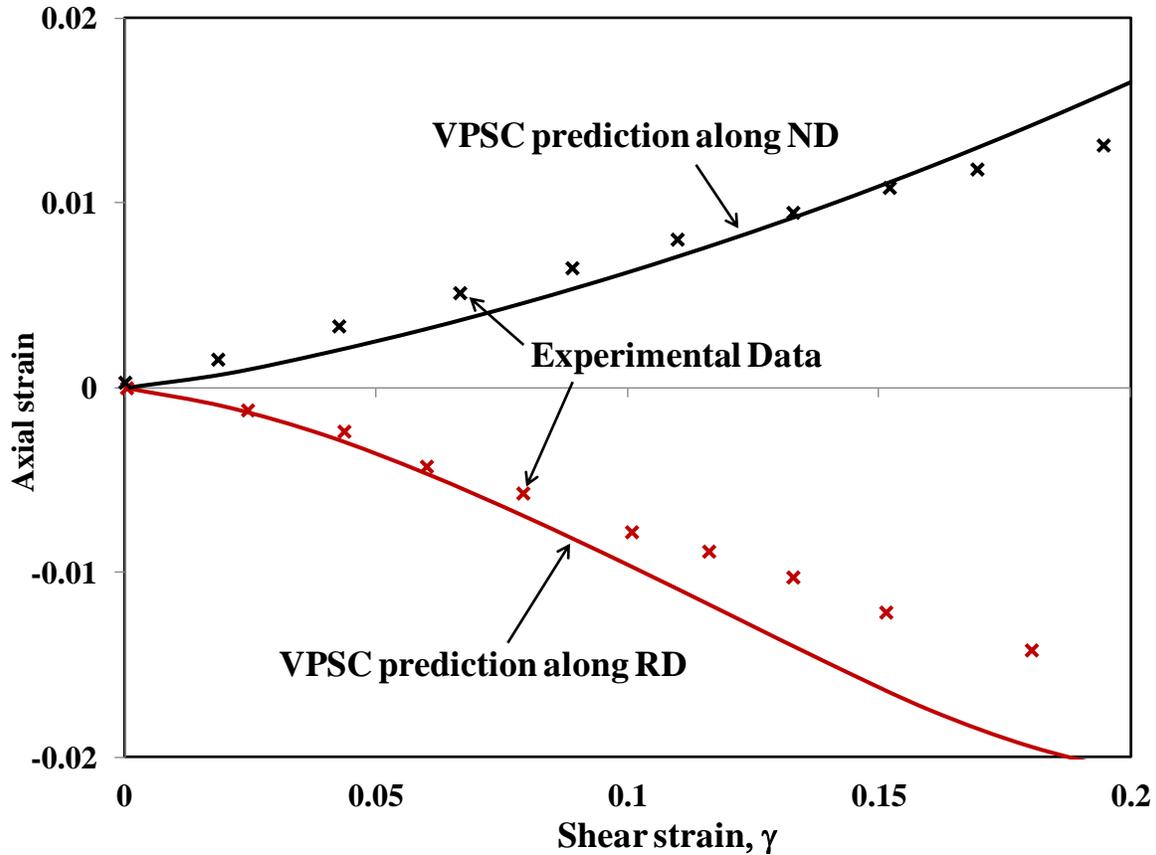

**Fig. 8:** Comparison between the experimental variation of the axial strain with the shear strain during room-temperature free-end torsion for a Mg AZ31 alloy reported in Guo et al. (2013) and prediction using the VPSC polycrystalline model assuming that all slip modes and tensile twinning are initially active for specimens with the long axis along RD and ND directions, respectively.

Before ending this section, we should remark that the identification of the VPSC model parameters was done using the mechanical data and microstructural information reported by Khan et al (2011) on an AZ31 sheet of 2 mm thickness. On the other hand, the torsional test results reported by Guo et al. (2013) were obtained from samples taken from an AZ31 magnesium plate of 76 mm thickness. Nevertheless, the agreement between VPSC model results and data for torsion about both RD and ND directions is very good.

## 3.2 Simulation of free-end torsion of AZ31 Mg using a macroscopic approach

The macroscopic model described by Eqs. (4-13) with yielding given by either the orthotropic form of the CPB06 criterion (Eq. 9) or by Hill (1948) (Eq. 13) will be also used to model the torsional response of AZ31 Mg. As already mentioned, Cazacu et al. (2006) criterion (see Eq. 9) accounts for both anisotropy and strength differential effects (tension-compression asymmetry). To identify the material parameters used in this criterion, the mechanical tests for uniaxial tension and compression reported in Khan et al. (2011) were used. To model the difference in hardening rates between tension and compression loadings observed experimentally, all these material parameters were considered to evolve with the accumulated plastic deformation. For $\bar{\varepsilon} \leq 0.05$, the anisotropy coefficients are almost constant. The numerical values of the model parameters corresponding to $\bar{\varepsilon} = 0.05$ and three other individual levels of equivalent plastic strains (up to 0.1 strain) are listed in Table 2, the values corresponding to any given level of plastic strain $\bar{\varepsilon}_p^j \leq \bar{\varepsilon} \leq \bar{\varepsilon}_p^{j+1}$ are obtained by linear interpolation, i.e.:

$$C_{ij}(\bar{\varepsilon}) = \alpha(\bar{\varepsilon}) C_{ij}(\bar{\varepsilon}_p^j) + (1-\alpha(\bar{\varepsilon})) C_{ij}(\bar{\varepsilon}_p^{j+1})$$
$$k(\bar{\varepsilon}) = \alpha(\bar{\varepsilon}) k(\bar{\varepsilon}_p^j) + (1-\alpha(\bar{\varepsilon})) k(\bar{\varepsilon}_p^{j+1})$$
(15)

The interpolation parameter α involved in Eq. (15) is defined as:

$$\alpha = \frac{\bar{\varepsilon} - \bar{\varepsilon}_p^j}{\bar{\varepsilon}_p^{j+1} - \bar{\varepsilon}_p^j}$$
(16)

Figure 9 shows the projection in the biaxial plane ($\sigma_{xx}$, $\sigma_{yy}$), with *x* being along RD and *y* being along TD, of the theoretical yield surfaces according to the orthotropic Cazacu et al. (2006) at different strain levels $\bar{\varepsilon}_p^j$, (up to 10%) along with the experimental data (symbols). for the given orthotropic AZ31 Mg alloy. The model correctly predicts that at initial yielding and below 8% strain, the tension-compression asymmetry is very pronounced (compare the tension-tension and compression-compression quadrants) while, for 8% strain and beyond, the difference in response between tension and compression becomes small, as observed experimentally (see the experimental stress-strain curves of Fig. 2).

| $\bar{\varepsilon}_p$ | $C_{22}$ | $C_{33}$ | $C_{12}$ | $C_{13}$ | $C_{23}$ | $C_{44}$ | $C_{55}$ | $C_{66}$ | $K$ |
|---|---|---|---|---|---|---|---|---|---|
| 0.05 | 1.090 | 3.342 | -0.168 | 0.098 | 0.243 | 0.730 | 7.30 | 7.74 | -0.625 |
| 0.06 | 1.072 | 2.905 | -0.595 | -0.279 | -0.096 | 1.039 | 10.2 | 11.02 | -0.520 |
| 0.08 | 1.099 | 1.439 | -0.817 | -0.516 | -0.350 | 1.128 | 11.21 | 11.95 | -0.215 |
| 0.10 | 1.082 | 0.885 | -0.762 | -0.657 | -0.509 | 1.058 | 10.12 | 11.21 | -0.169 |

**Table** 2: Material parameters (orthotropy coefficients $C_{ij}$ and strength-differential parameter *k*) involved in Cazacu et al. (2006) yield criterion for AZ31 Mg alloy corresponding to different values of plastic strain, $\bar{\varepsilon}^p$; for any strain level $C_{11}$ is set to unity.

Since according to Hill (1948) criterion the mechanical response is the same in tension and compression, only the experimental flow stress data in uniaxial tension were used for the identification of the material parameters. The numerical values of all the anisotropy coefficients involved in Hill (1948) criterion (see Eq. 14) are given in Table 3.

| $\bar{\varepsilon}_p$ | F | G | H | N | M | L |
|---|---|---|---|---|---|---|
| 0.03 | 0.195 | 0.276 | 0.724 | 2.09 | 11.186 | 12.43 |
| 0.05 | 0.206 | 0.288 | 0.712 | 2.167 | 11.813 | 15.245 |
| 0.06 | 0.217 | 0.297 | 0.703 | 2.184 | 12.184 | 15.873 |
| 0.08 | 0.228 | 0.298 | 0.702 | 2.165 | 12.210 | 16.136 |
| 0.10 | 0.246 | 0.312 | 0.688 | 2.167 | 11.816 | 15.725 |

**Table** 3: Orthotropy coefficients involved in Hill's (1948) yield criterion (Eq. 13) for AZ31 magnesium alloy corresponding to different values of plastic strain.

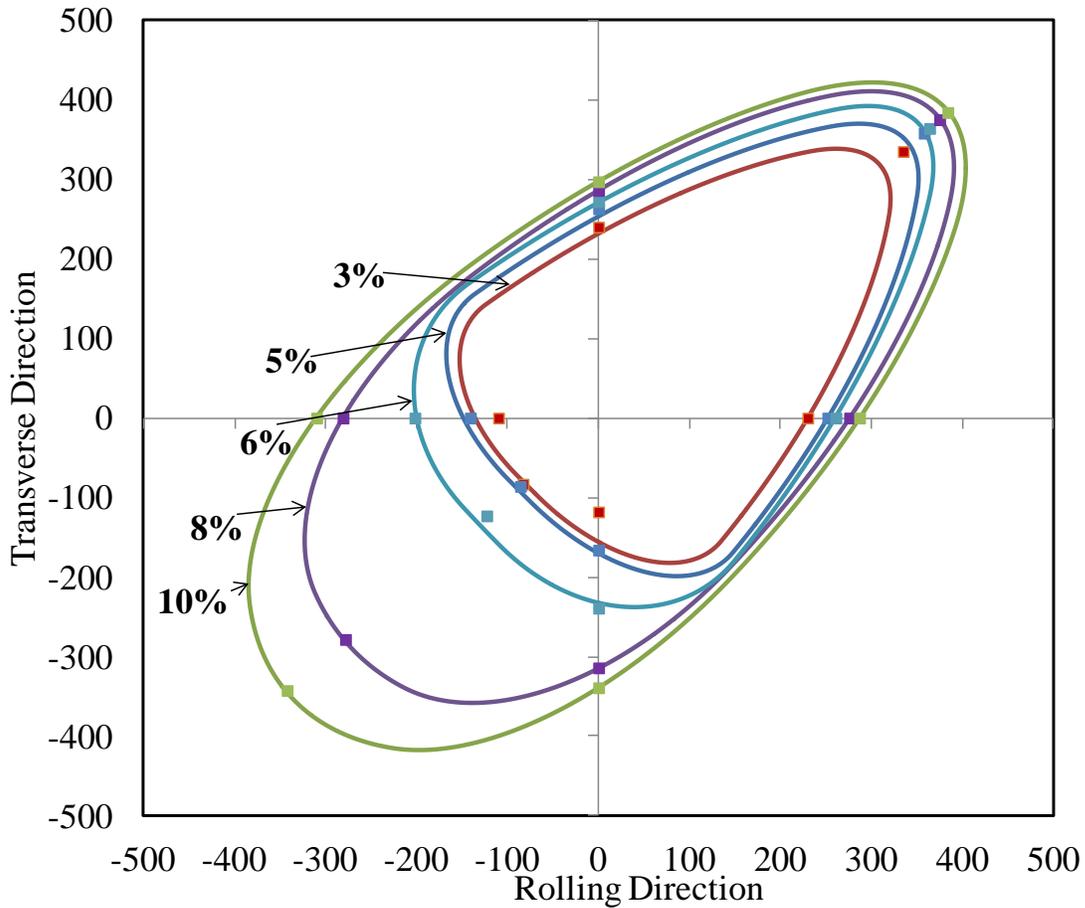

**Fig. 9:** Comparison between the theoretical yield surfaces, according to the orthotropic Cazacu et al. (2006) criterion, corresponding to different levels of accumulated plastic strain and data for an AZ31 Mg alloy reported by Khan et al. (2011). Note the evolution of the shape of the yield locus from a triangular shape at lower strains to an elliptical shape at larger strains.

The yield surfaces associated with Hill (1948) yield criterion corresponding to different levels of equivalent plastic strains are plotted in Fig. 10. Note that the shape of Hill's (1948) yield surface is always elliptical and does not capture the tension-compression asymmetry of the material.

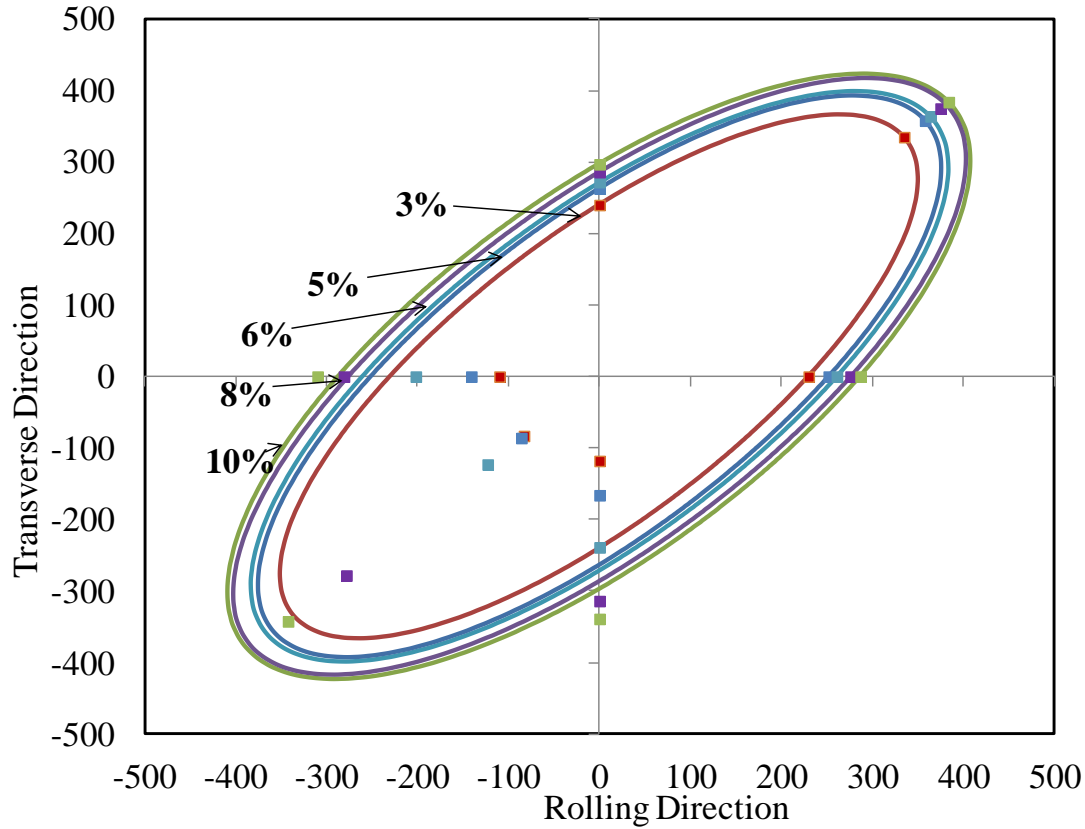

**Fig**. 10: Yield surfaces of AZ31 Mg corresponding to fixed levels of accumulated plastic strain according to Hill (1948) orthotropic criterion against mechanical test data (symbols). Stresses are in MPa. Data from Khan et al. (2011).

### 3.3 FE simulation of the torsional response of AZ31 Mg using the macroscopic models

For anisotropic materials, the boundary value problem associated to torsional loading cannot be solved analytically. Hence, were solved it numerically using the finite-element (FE) method. All the simulations were carried out with the commercial FE code ABAQUS, using user material routines (UMAT) that we developed for the anisotropic elastic-plastic model with yielding described by Cazacu et al. (2006) criterion and Hill (1948) criterion, respectively. A fully implicit integration algorithm was used for solving the governing equations. The geometry of the specimen and FE mesh used in all the calculations is shown in Fig. 11. The FE mesh consists of 1290 hexahedral elements with reduced integration (ABAQUS C3D8R). The initial minimal section was meshed with 10 layers of elements and three elements were used along the wall thickness. The usual definitions of the axial and shear strains are used, namely:

$$\varepsilon = \ln\left(1 + \frac{u}{L_0}\right) \quad \text{and} \quad \gamma = \frac{\Phi r}{L_0}, \tag{17}$$

where $r$ is the current radius, $L_0$ is the initial length, $u$ is the axial displacement, and $\Phi$ is the twist angle. In all the FE simulations, equal-sized time increments $\Delta t = 10^{-3}$ s were considered. Between five and six iterations per increment were necessary for convergence in the return mapping algorithm, the tolerance in satisfying the yield criterion being $10^{-7}$ (0.1 Pa). The Young modulus and Poisson coefficient used are E=45 GPa and $\nu$=0.3, respectively. The parameters involved in the isotropic hardening law (Eq. 8) was identified from the uniaxial tension stress-strain response along rolling direction, the numerical values being $A_0 = 315.4$ MPa, $A_1 = 140.6$ MPa, $A_2 = 16.3$.

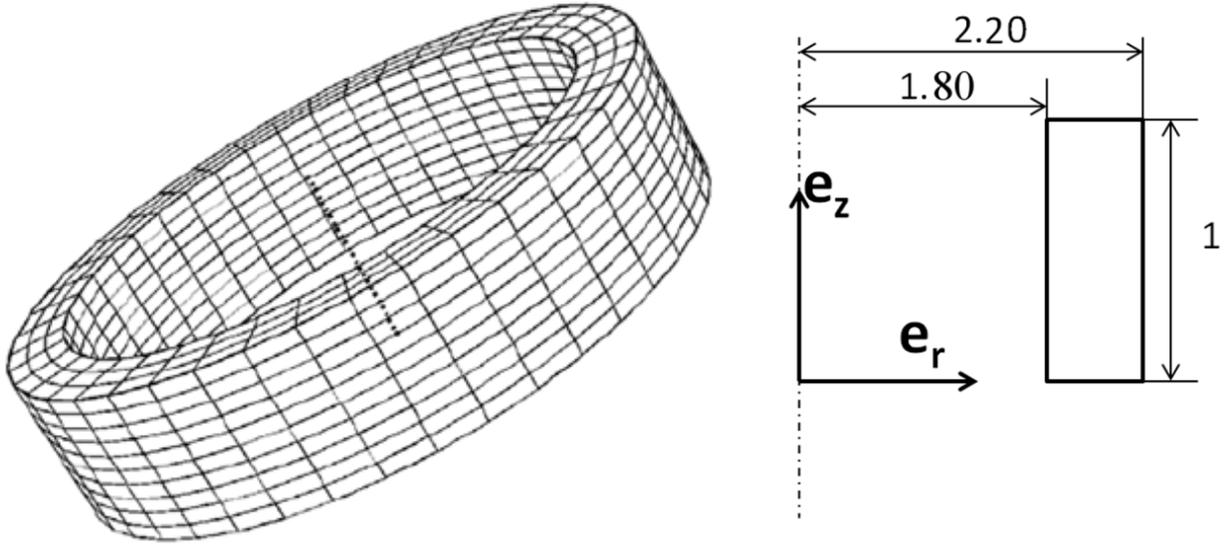

**Fig. 11**. Sample geometry and dimensions (mm) and finite-element mesh.

Comparisons between FE simulations using the macroscopic models with yielding according to Cazacu et al. (2006) criterion and Hill (1948), respectively in conjunction with the isotropic hardening law of Eq. (8) and the data for Mg AZ31 under free-end torsion along the RD and ND directions are presented in Figs. 12 and 13, respectively. Note that both models predict that the specimen contracts axially in the RD direction. However, only using Cazacu et al. (2006) criterion (which accounts for anisotropy and tension-compression asymmetry in yielding) both the level of axial strains and the slope of the axial vs. shear strain curve are accurately predicted. On the other hand, Hill (1948) criterion largely underestimates the axial strains that develop in RD (see Fig. 13a) and cannot capture any Swift effects in ND torsion (see Fig. 13b). It is also worth noting that the axial effects predicted by Hill's (1948) criterion for torsion of the RD specimen are very close with the VPSC polycrystal model predictions obtained when twinning activity is neglected (see Fig. 13a). On the other hand, Cazacu et al. (2006) yield criterion in conjunction with the same isotropic hardening law captures well the experimental data in RD and ND directions, and the predictions are very close to that obtained using the VPSC polycrystal model when all slip modes and tensile twinning are considered operational (see Fig. 12 and Fig. 14, respectively).

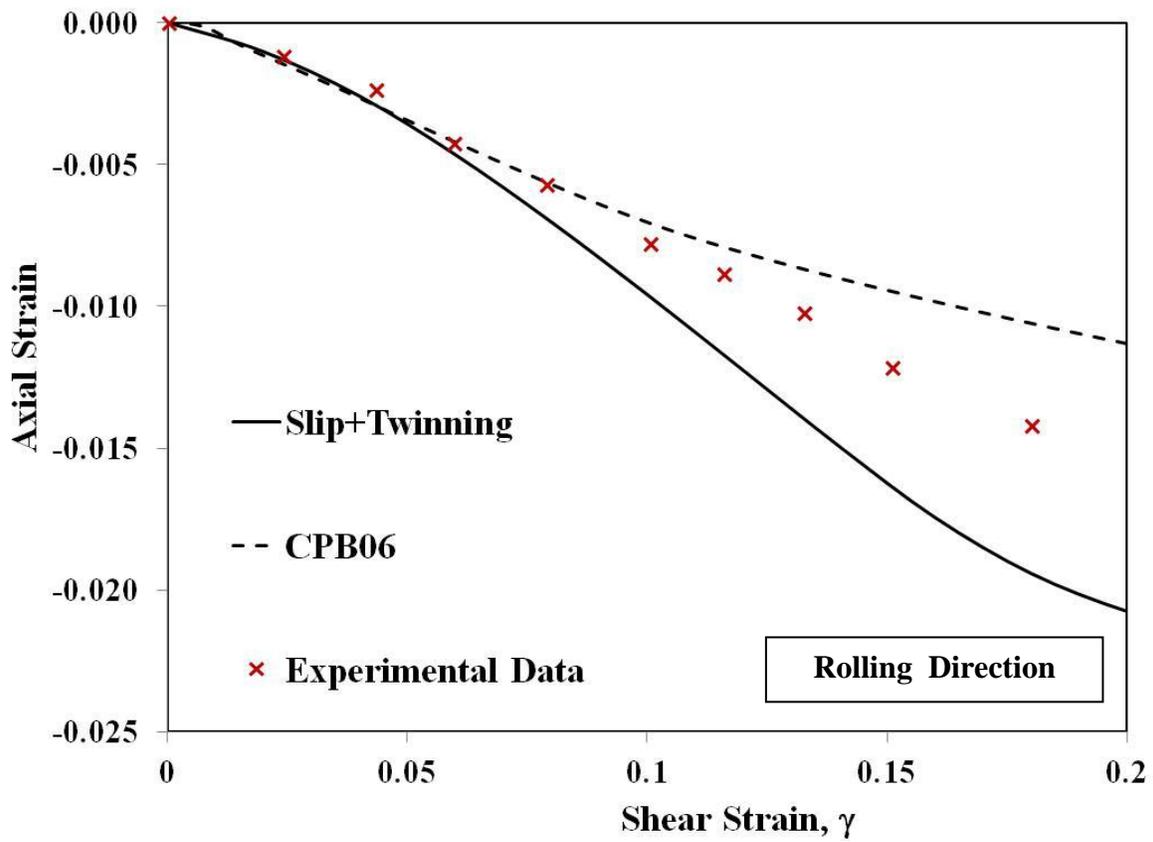

(a)

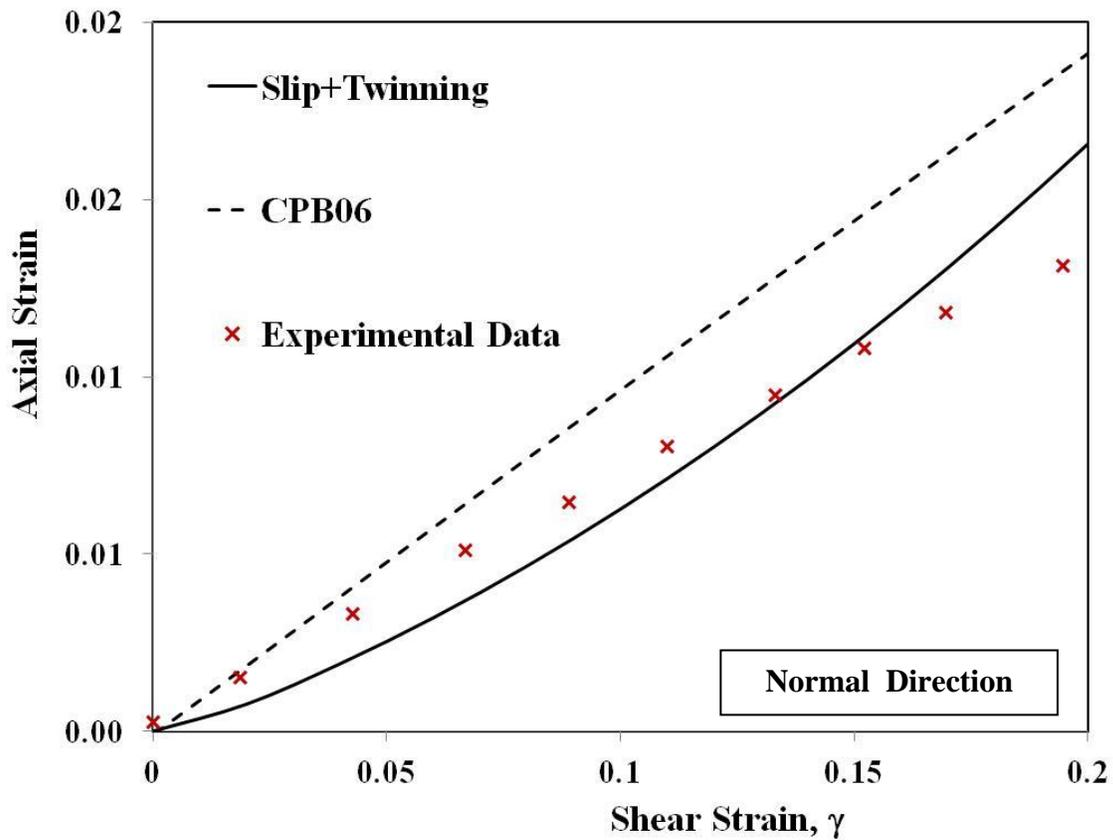

(b)

**Fig. 12:** Comparison of the experimental variation of the axial strain with the shear strain (symbols) during room-temperature free-end torsion of AZ31 magnesium alloy (data from Guo et al. (2013)) with the predictions according to the orthotropic Cazacu et al. (2006) criterion, and the polycrystal model simulations considering that all slip modes and twinning are active: (a) RD torsion; (b) ND torsion.

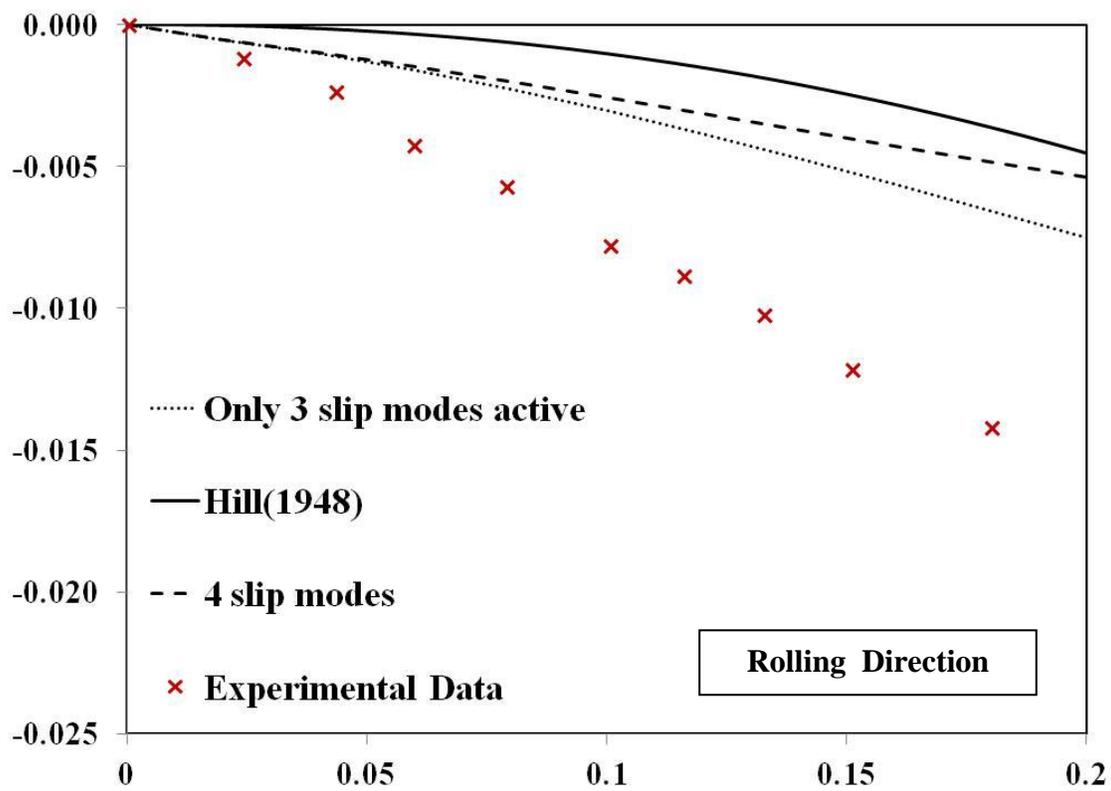

(a)

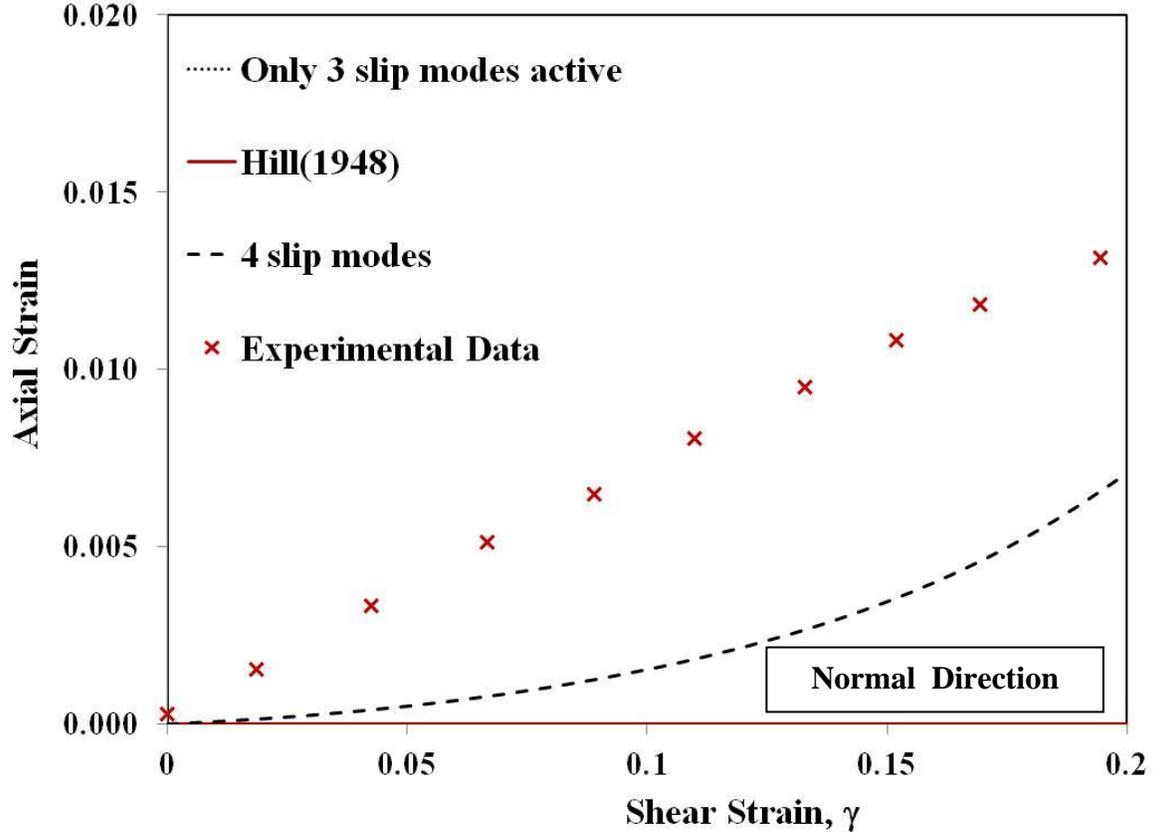

(b)

**Fig. 13:** Comparison of the experimental variation of the axial strain with the shear strain (symbols) during room-temperature free-end torsion of AZ31 magnesium alloy (data from Guo et al. (2013)) with the predictions according to the orthotropic Hill's (1948) criterion, and the VPSC polycrystal model simulations considering that all slip modes and twinning are active: (a) RD torsion; (b) ND torsion.

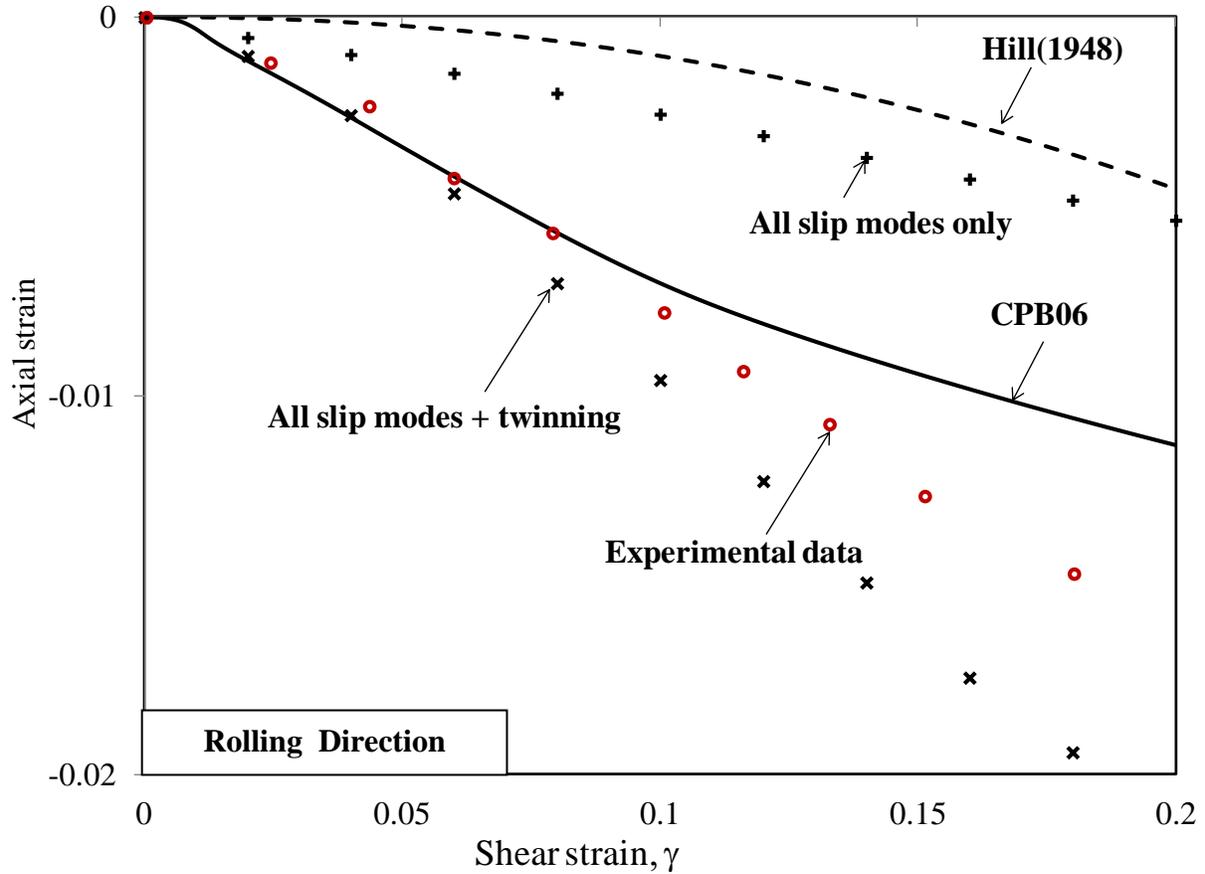

**Fig.14**: Comparison of the experimental variation of the axial strain with the shear strain (symbols) during free-end torsion of AZ31 magnesium alloy specimens with long axis along the rolling direction (RD) with the predictions according to the orthotropic Cazacu et al. (2006) criterion, Hill (1948), and the VPSC simulations considering that all slip modes and twinning are active (x), and when twinning activity is neglected (+). Data after Guo et al. (2013).

In summary, these results show for the first time that unless tension-compression asymmetry in plastic flow is captured at every scale, the peculiarities of the torsional response of polycrystalline Mg AZ31 cannot be predicted. Moreover, it is confirmed the correlation between tension-compression asymmetry and Swift effects that was established based on the predictions of Cazacu et al. (2006) macroscopic model (see Revil-Baudard et al., 2014). It was also demonstrated that only by considering that all slip modes (basal, prismatic and pyramidal <c+a>, and pyramidal <a>) and tensile twinning are active, both the mechanical behavior and the microstructure evolution of the Mg AZ31 alloy in shear can be predicted with accuracy.

## 4. Conclusions

In this paper, the understanding of Swift effect in Mg AZ31 has been improved at different scales, using macroscopic modeling and the crystal plasticity framework. Independent of the modeling framework used, it was demonstrated the correlation between the tension-compression asymmetry of the material and the sign and level of the axial strains that develop during torsion in Mg AZ31 alloy. Comparison between the prediction of axial strain vs. shear strain for torsion in the RD direction obtained using the VPSC model (with twinning considered active), and Cazacu et al. (2006) yield criterion, in conjunction with appropriate evolution laws for the anisotropy coefficients and the strength differential parameter *k*, and the experimental data from Guo et al. (2013) show that both models capture the unusual behavior in torsion of Mg AZ31 with great accuracy. Moreover, the predictions obtained with the macroscopic model and the VPSC model are very close. This reinforces the fact that only by accounting for tension-compression asymmetry in the plastic flow, axial effects can be predicted. Therefore, both approaches point to the correlation between tension-compression asymmetry and Swift effects in AZ31 Mg. On the other hand, Hill's (1948) yield criterion, which doesn't account for the tension-compression asymmetry of the material, cannot capture the Swift effect in the ND direction and largely underestimates these effect in RD direction, the level of axial strains predicted being very close to that obtained using VPSC when twinning neglected.

## References

1. ABAQUS. User's Manual for Version 6.8. Volumes I–V. Dassault Systemes Simulia Corp., Providence, RI, 2009.
2. Barnett, M.R. 2007a. Twinning and the ductility of magnesium alloys. Part I: Tension twins. Material Science and Engineering A 464, 1-7.
3. Barnett, M.R. 2007b. Twinning and the ductility of magnesium alloys. Part II: Contraction twins. Material Science and Engineering A 464, 8-16.
4. Agnew, S.R., Duygulu, O., 2005. Plastic Anisotropy and the role of non-basal slip in magnesium alloy AZ31B. Int. J. Plasticity 21, 1161-1193.

5. Beausir, B., Toth, L. S., Qods, F., Neale, K. W., 2009. Texture and mechanical behavior of Magnesium during free-end torsion. J. Eng. Mat. Technol. 131, 1-15.
6. Billington, E., 1977a. Non-linear mechanical response of various metals: I Dynamic and static response to simple compression, tension and torsion in the as received and annealed states. J. Phys. D 10, 519-531.
7. Billington, E., 1977b. Non-linear mechanical response of various metals: II permanent length changes in twisted tubes. J. Phys. D 10, 533-552.
8. Billington, E., 1977c. Non-linear mechanical response of various metals: III Swift effect considered in relation to the stress-strain behaviour in simple compression, tension and torsion. J. Phys. D 10, 553-569.
9. Cazacu, O., Revil-Baudard, B., Barlat, F., 2013. New interpretation of monotonic Swift effects: Role of tension–compression asymmetry. Mechanics of Materials 57, 42–52.
10. Cazacu, O., Plunkett, B., Barlat, F., 2006. Orthotropic yield criterion for hexagonal closed packed materials. Int. J. Plasticity 22, 1171-1194.
11. Duchene, L., Houdaigui, F.E., Habraken, A.M., 2007. Length changes and texture prediction during free-end torsion test of copper bars with FEM and remeshing techniques. Int. J. Plasticity 19, 1417-1438.
12. Green, A. E. and Naghdi, P.M., 1965. A general theory of an elastic-plastic continuum. Archive Rational Mech. & Anal., 18, 251-281.
13. Guo X.Q.,Wu W., Wu P.D., Qiao H., An K., Liaw P.K. 2013. On the Swift effect and twinning in a rolled magnesium alloy under free-end torsion. Scripta Mater. 69, 319–322.
14. Habraken, A.M., Duchene, L., 2004. Anisotropic elasto-plastic finite element analysis using a stress–strain interpolation method based on a polycrystalline model. Int. J. Plasticity 20, 1525–1560.
15. Hill, R., 1948. A theory of the yielding and plastic flow of anisotropic materials. Proc. Royal. Soc. London A 193, 281-297.
16. Jiang, L., Jonas, J.J., Boyle, K., Martin, P., 2008. Deformation behavior of two Mg alloys during ring hoop tension testing. Material Science and Engineering A 492, 68-73.
17. Jain, A., Agnew, S.R., 2007. Modeling the temperature dependent effect of twinning on the behavior of magnesium alloy AZ31B sheet. Material Science and Engineering A 462, 29-36.

18. Lebensohn, R. A., Tome, C. N., 1993. A self-consistent anisotropic approach for the simulation of plastic deformation and texture development of polycrystals: Application to zirconium alloys. Acta Metall. Mater. 41, 2611-2624.
19. Lou, X.Y., Li, M., Boger, R.K., Agnew, S.R., Wagoner, R.H., 2007. Hardening evolution of AZ31B Mg sheet. Int. J. Plasticity 23, 44-86.
20. Khan, A., Pandey, A., Gnaupel-Herold, T., Mishra, R.K., 2011. Mechanical response and texture evolution of AZ31 alloy at large strains for different strain rates and temperatures. Int. J. Plasticity 27, 688-706.
21. Mekonen M.N., Steglich D., Bohlen J., Letzig D., Mosler J. 2012. Mechanical characterization and constitutive modeling of Mg alloy sheets. Materials Science and Engineering A 540, 174–186.
22. Plunkett, B., Lebensohn, R.A., Cazacu, O., Barlat, F., 2006. Evolving yield function of hexagonal materials taking into account texture development and anisotropic hardening. Acta Mater. 54, 4159–4169.
23. Plunkett, P., Cazacu, O., Lebensohn, R.A., Barlat, F., 2008. Elastic-viscoplastic anisotropic modelling of textured metals and validation using the Taylor cylinder impact test. Int. J. Plasticity 23, 1001-1021.
24. Proust, G., Tomé, C.N., Jain, A., Agnew, S.R., 2009. Modeling the effect of twinning and detwinning during strain-path changes of magnesium alloy AZ31. Int. J. Plasticity, 25, 861-880.
25. Revil-Baudard, B., Chandola, N., Cazacu, O., Barlat, F., 2014. Correlation between tension-compression asymmetry and Swift effects in various polycrystalline materials. J. Mech. Phys. Solids (accepted for publication).
26. Swift, H., 1947. Length changes in metals under torsional overstrain. Engineering 163, 253-257
27. Tomé, C.N., Lebensohn, R.A., Kocks U.F., 1991. A model for texture development dominated by deformation twinning: application to Zirconium alloys. Acta Metall. Mater. 39, 2667-2680.
28. Tomé, C.N., Lebensohn, R., 2004. "Self-consistent homogeneization methods for texture and anisotropy" in: *Continuum Scale Simulation of Engineering Materials:*

*Fundamentals, Microstructures*, Process Applications. D. Raabe, F. Roters, F. Barlat and L.-Q. Chen (Eds.), Willey, pp. 352-378 (2004).

29. Toth, L.S., Jonas, J. J., Gilormini, P. Bacroix, B., 1990. Length changes during free end torsion: a rate sensitive analysis. Int. J. Plasticity 6, 83-108.
30. Wang, H., Raeisinia, B., Wu, P.D., Agnew S.R., Tomé, C.N., 2010. Evaluation of self-consistent polycrystal plasticity models for magnesium alloy AZ31B sheet. International Journal of Solids and Structures 47, 2905-2917.